\documentclass[a4paper]{article}

\usepackage{INTERSPEECH2021}

\usepackage{amsmath,graphicx}
\usepackage{booktabs}
\usepackage{amssymb}
\usepackage{cite}
\usepackage[urlcolor=blue]{hyperref}

\title{Adversarially Learning Disentangled Speech Representations for Robust Multi-factor Voice Conversion}
\name{Jie Wang$^{1, \dagger}$,
\thanks{$\dagger$ Work performed while interning at Huya Inc.}
        Jingbei Li$^1$,
        Xintao Zhao$^{1, \dagger}$,
       Zhiyong Wu$^{1,3,\ddagger}$,\thanks{$\ddagger$ Corresponding author.} 
       Shiyin Kang$^2$,
       Helen Meng$^{1,3}$}

\address{
    $^1$ Shenzhen International Graduate School, Tsinghua University, Shenzhen, China\\
$^2$ Huya Inc., Guangzhou, China\\
$^3$ The Chinese University of Hong Kong, Hong Kong SAR, China
    }
\email{   \small{\{jie-wang19,lijb19,zxt20\}@mails.tsinghua.edu.cn,
    kangshiyin@huya.com
	\{zywu,	hmmeng\}@se.cuhk.edu.hk}}

\begin{document}



\maketitle

\begin{abstract}
Factorizing speech as disentangled speech representations
is vital to achieve highly controllable style transfer in voice conversion (VC).
Conventional speech representation learning methods in VC only factorize
speech as speaker and content, lacking controllability on other prosody-related factors. 
State-of-the-art speech representation learning methods for more speech factors 
are using primary disentangle algorithms such as
random resampling and 
ad-hoc bottleneck layer size adjustment,
which however is hard to ensure robust speech representation disentanglement.
To increase the robustness of highly controllable
style transfer on multiple factors in VC,
we propose a disentangled speech representation learning framework based on 
adversarial learning.
Four speech representations characterizing 
content, timbre, rhythm and pitch are extracted,
and further disentangled by an adversarial Mask-And-Predict (MAP)  network inspired by BERT.
The adversarial network is used to minimize the correlations between the speech representations, by randomly masking and predicting one of the representations from the others.
Experimental results show that the proposed framework significantly improves the robustness of VC on multiple factors by increasing the speech quality MOS from 2.79 to 3.30 and decreasing the MCD from 3.89 to 3.58.
\end{abstract}
%
\noindent\textbf{Index Terms}: disentangled speech representation learning, 
multi-factor voice conversion, 
prosody control,
adversarial learning,
gradient reverse layer
\section{Introduction}
\label{sec:intro}

Voice conversion (VC) aims at converting the input speech of a source speaker to sound as if uttered by a target speaker without altering the linguistic content \cite{kain1998spectral}.
Besides the conversion of timbre, the conversion can also be conducted in various domains such as 
pitch, rhythm or other non-linguistic domains. 
Representation learning methods for these speech factors have already been proposed and applied in
many research fields in speech processing
\cite{sadhu2021wav2vec, niizumi2021byol, liu2020mockingjay}.
However, directly applying the speech representations extracted by these methods in VC may cause unexpected conversions of other speech factors as they may be not necessarily orthogonal.
Therefore, disentangling the representations of intermingling various informative factors in speech signal is crucial to achieve highly controllable VC \cite{li2018deep}, few-shot synthesis \cite{van2020vector} and speaker adaptation \cite{wang2020spoken}.

Ideally, the VC technology is able to preserve the linguistic information and convert para-linguistic information.
Conventionally, only speaker and content information are factorized in VC.
Prosody, the important cue in speech signals, is not properly modeled in the VC framework.
There are explorations on the controllability of prosody in VC, among which SpeechSplit \cite{qian2020unsupervised} is noticeable for its high controllability on multi speech factors. 
However, information-constraining bottleneck encoding layers can only gain limited disentanglement.
The entanglement of perceptual attributes engenders the low similarity which is elaborated in Section \ref{sec:related}.

In this paper, to achieve highly controllable style transfer for multiple factors VC, we propose a disentangled speech representation learning framework based on adversarial learning.
The proposed framework explicitly removes the correlations between the speech representations which characterize different factors of speech by an adversarial network inspired by BERT \cite{devlin2018bert}. 
The speech is firstly decomposed into four speech representations which represent content, timbre and other two prosody-related factors, rhythm and pitch.
During training, one of the speech representations will be randomly masked and inferred from the remaining representations by the adversarial MAP
network. 
The MAP network is trained to maximize the correlations between the masked and the remaining representations, while the speech representation encoders are trained to minimize the correlations by taking the reversed gradient of the MAP network. 
In this way, the representation learning framework is trained in the adversarial manner, with speech representation encoders trying to disentangle the representations while MAP network trying to maximize the representation correlations.
The decoder reconstructs the speech from the representations during training and achieves VC on multiple factors by replacing the corresponding speech representations.
Experimental  results  show  that  the  proposed  speech  representation learning  framework  significantly  improves  the robustness of VC on multiple factors, decreasing the MCD from 3.89 to 3.58 and outperforms state-of-the-art speech representation learning methods for multiple factors VC by a gap of 0.51 speech quality MOS.

\section{Related Work}
\label{sec:related}
Prosody is an important component of speech which usually reflects rhythm, intonation etc and there are explorations on prosody transfer as expressive and controllable speech synthesis is attaining more attention\cite{skerry2018towards, 9420276}.
A combination of explicit and latent variables are adopted to achieve high controllable and natural speech synthesis
\cite{valle2020mellotron}.
The explicit variables contain pitch contour, loudness besides speaker embedding.
The latent variables contain rhythm and duration information etc which are obtained from reference encoder \cite{skerry2018towards}, denoted by global style tokens \cite{wang2018style} or enhanced by pre-trained language model \cite{zhao2020enhancing}.


Conventionally, only speaker and content information are factorized in VC.
Unsupervised learning based-methods are garnering attention for the advantage of no need for text transcriptions and
quite a lot of
them are based on auto-encoder architecture.
Variational autoencoder \cite{huang2018voice,elgaar2020multi}, vector quantization 
\cite{wu2020one} 
and instance normalization based methods \cite{chou2019one,chen2020again} were proposed to better model the latent space and pursue the regularization property.
Previous studies of prosody conversion mainly focus on transformation of F0 related features
\cite{du2020spectrum,lian2021towards} 
which gains limited conversion similarity.

The entanglement between different speech representations causes the low similarity and naturalness of synthesized speech whether in prosody transfer or timbre transfer.
Expressive and high controllable speech synthesis systems share the same principle of disentangling multiple speech factors like speaker, linguistic and prosody-related information.
In order to foster disentanglement,
adversarial training \cite{chou2018multi, ocal2019adversarially}, contrastive learning \cite{ebbers2020adversarial, li2020cvc,van2020vector}, and mutual information minimization \cite{chen2016infogan, cheng2020improving} are applied to attenuate the information leakage.
However, only the disentanglement between two factors, e.g., style, content or speaker
\cite{yuan2021improving,williams2019disentangling}
are taken into account.


\begin{figure}[t]
  \centering
  \includegraphics[width=0.95\linewidth]{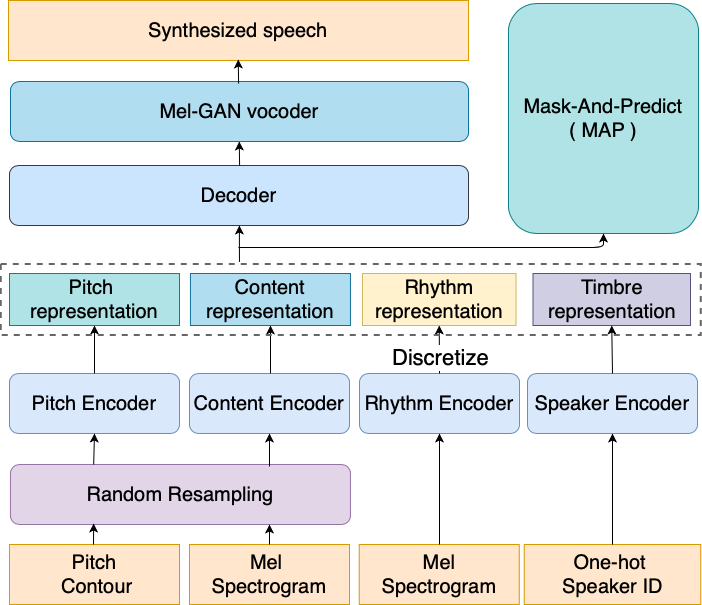}
  \caption{\textit{Overall architecture. }}
  \label{fig:structure}
\end{figure}

Effective disentanglement modeling for multi-factor voice conversion remains a challenging problem.
To overcome 
that prosody is also converted while transferring timbre
in conventional VC,
different information bottlenecks are applied to
decompose the speaker information into timbre and other prosody-related factors such as
rhythm and pitch \cite{qian2020unsupervised, takahashi2021hierarchical}.
To improve disentanglement, restricted sizes of bottleneck layers encourage the encoders to discard the information which can be learnt from other bottlenecks.
Random resampling 
\cite{qian2020unsupervised}
is also proposed to use in the information bottlenecks to remove rhythm information from content and pitch representations.

However, without explicit disentanglement modeling, random resampling and restricting the sizes of bottleneck layers can only gain limited disentanglement of speech representations.
Random resampling which is implemented as dividing and resampling speech segment using linear interpolation along time dimension can only be used in removing time-related information such as rhythm.
Moreover, random resampling is proved as a partial disentanglement algorithm that only contaminate a random portion of rhythm information \cite{qian2020unsupervised}.
The content encoder actually is a residual encoder
which cannot ensure that the content information is only modeled in the content representation.

\section{Methodology}
\label{sec:method}

We aim to improve the converted speech quality by explicitly disentangling speech representations.
As shown in Figure \ref{fig:structure}, the architecture of the proposed disentangled speech representation learning framework is composed of three sub-networks: (i) multiple speech representation encoders which encode speech into different speech representations characterising content, timbre, rhythm and pitch,
(ii) 
the MAP network that is trained to capture the correlations between different speech representations based on mask-and-predict operations,
(iii) a decoder which is employed to generate spectrogram from these disentangled speech representations.
\begin{figure}[t]
  \centering
  \includegraphics[width=0.85\linewidth]{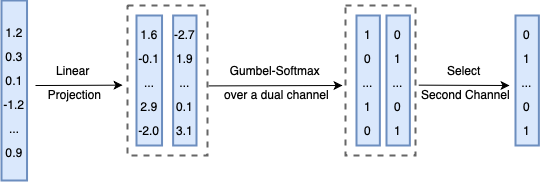}
  \caption{\textit{Multi-label Binary Vectors (MBV) are learned as the discrete rhythm representation.}}
  \label{fig:mbv}
\end{figure}
Afterwards, the neural vocoder is utilized to synthesize audio from the generated mel spectrogram.
\subsection{Speech representation extraction}
Content, rhythm and pitch encoders of \cite{qian2020unsupervised} are adopted to extract content, rhythm and pitch representations from mel spectrogram and pitch contour respectively at frame-level, as shown in Figure \ref{fig:structure}.
Different from \cite{qian2020unsupervised}, one-hot speaker labels (ID) are encoded by a training embedding table to obtain the timbre representation in the proposed framework.

Besides the random resampling and restriction on sizes of bottleneck layers which is adopted in \cite{qian2020unsupervised} and analyzed in Section \ref{sec:related} to gain the limited disentanglement, we put on stricter constrain on the rhythm encoder output.
As shown in Figure \ref{fig:structure}, the output of rhythm encoder is discretized.
The input of the rhythm encoder is mel spectrogram which contains much information
and there is a distinct possibility that various messy information will be encoded into the rhythm representation with merely random resampling and restriction on sizes of bottleneck layers.
Considering that discrete variable typically offers a substantially reduced model capacity,
we applied the  
Multilabel-Binary Vectors (MBV) \cite{liu2019unsupervised} 
as the discrete rhythm representation.
The MBV with Gumbel-Softmax demonstrated in Figure \ref{fig:mbv} is more data efficient than one-hot vector and continuous variable in information distillation task \cite{liu2019unsupervised}.

\subsection{Adversarial learning for speech representation disentanglement}
\begin{figure}[t]
  \centering
  \includegraphics[width=7.05cm,height=7.25cm]{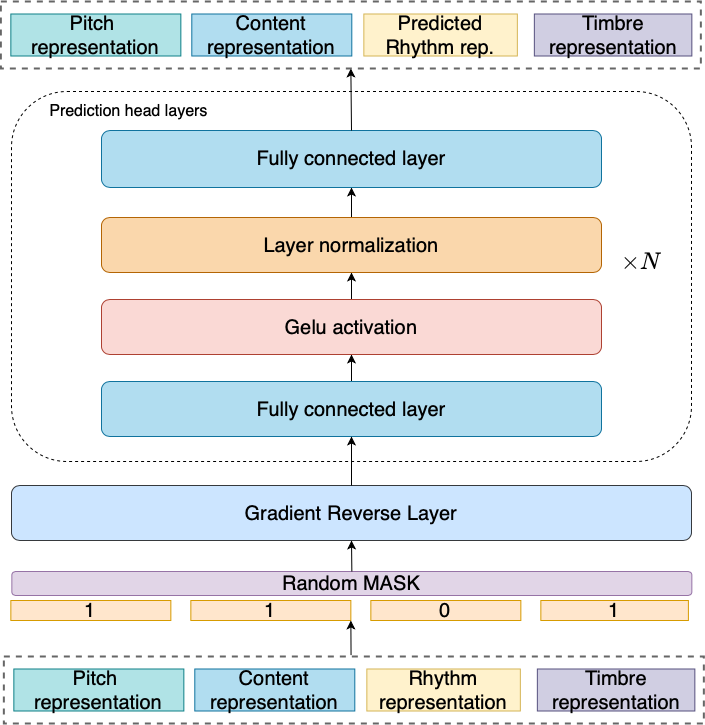}
  \caption{\textit{Architecture of Mask-And-Predict (MAP) network.}}
  \label{fig:map}
\end{figure}

To address the limitations of SpeechSplit \cite{qian2020unsupervised} as discussed in Section \ref{sec:related}, an adversarial Mask-And-Predict (MAP) network
inspired by BERT \cite{devlin2018bert} is designed to explicitly disentangle the extracted speech representations.
During training, one of the four speech representations is randomly masked and the adversarial network tries to infer the masked representation from the other three representations.
The adversarial network is composed of a gradient reverse layer (GRL) \cite{ganin2016domain} and a stack of prediction head layers \cite{liu2020mockingjay} which has been used in masked acoustic modeling. 
As delineated in Figure \ref{fig:map}, in the MAP module,
each prediction head layer is composed of a fully-connected layer, GeLU activation,
layer normalization
and another fully-connected layer.
The gradient of the adversarial network is reversed by GRL
before backward propagated to the speech representation encoders.
$L1$ loss is adopted here to measure the adversarial loss which is
demonstrated below:
\begin{equation}
Z = (Z_r, Z_c , Z_f, Z_u)
\end{equation}
\begin{equation}
M \in \left\{(0, 1, 1, 1), (1, 0, 1, 1), (1, 1, 0, 1), (1, 1, 1, 0)\right\}
\end{equation}
\begin{equation}
\small
\label{eq:adv}
{L}_{adversarial} = 
\left|\left|(1-M)\odot (Z - {\rm{MAP}} (M\odot Z))
\right|\right|
\end{equation}
where $\odot$ is element-wise product operation,
$L_{adversarial}$ is adversarial loss,
$Z$ is the concatenation of $Z_r$, $Z_c$, $Z_f$, $Z_u$ denoting rhythm, content, pitch and timbre representations respectively,
$M$ is a randomly selected binary mask corresponding to the dropped region with a value of 0 
and 1 for unmasked representations.
The MAP network is trained to predict the masked representation as accurate as possible by minimizing the adversarial loss, 
while in the backward propagation, the gradient is reversed which encourages the representations learned by the encoder contain as little mutual information as possible.

\subsection{VC with disentangled speech representations}

During training, four speech representations are extracted from same utterance and decoder is trained to reconstruct mel spectrogram from speech representations with the loss defined as:
\begin{equation}
\small
\label{eq:recon}
{L}_{reconstruct} = \|S-\hat{S}\|_{2}^{2}
\end{equation}
where $S$ and $\hat S$ are the mel spectrograms of the input and reconstructed speech respectively.
Final objective function with trade-off parameters is given in:
\begin{equation}
\small
\label{eq:loss}
{Loss} = \alpha * L_{adversarial}
+ \beta * L_{reconstruct}
\end{equation}
where $\alpha$, $\beta$
are the loss weights for adversarial loss and reconstruction loss.
To improve the robustness of proposed framework, the loss weight for the reconstruction loss is designed to be exponential decaying.
MelGAN \cite{kumar2019melgan} is adopted as the vocoder
for the high fidelity speech and fast decoding speed.


\section{Experiment}
\label{sec:typestyle}
\subsection{Training setup}
The experiments were performed on the CSTR VCTK corpus \cite{veaux2016superseded}.
We randomly selected 107 speakers including 
62 females and 45 males.
After pre-processing, the corpus duration for experiment was 
43.5 hours in total, 35.3 hours for training and 4.1 hours validation and test respectively.
All the audios were down-sampled to 16000Hz. 
Mel spectrograms were computed through a short time Fourier transform (STFT) using a 50 ms frame size, 12.5 ms frame hop, and a Hann window function.
We transformed the STFT magnitude to the mel scale using an 80 channel mel filterbank spanning 125 Hz to 7.6 kHz, followed by log dynamic range compression.
The filterbank output magnitudes were clipped to a minimum value of 0.01.
The weights of adversarial loss was fixed to $10^{-1}$.
The weight of reconstruction loss $\beta$ applied an initial weight of $1$ with decay factor of 0.9 every 500,000 steps.
We trained a vanilla SpeechSplit \cite{qian2020unsupervised} as the \textbf{Baseline} system and the system described in Section \ref{sec:method}
as \textbf{Proposed}.
We used a pretrained Mel-GAN vocoder on VCTK corpus to synthesize the audios from the spectrogram.
There were three factors involved in the conversion process and
we conducted seven types conversion including rhythm-only conversion,
pitch-only conversion,
timbre-only conversion 
and combinations of them.
We evaluated baseline and proposed systems under the same settings, otherwise mentioned.
We programmed all neural networks used in the experiments based on an open source pytorch implemention of SpeechSplit
\cite{qian2020unsupervised}.
We trained all models with a batch size of 64 for 800,000 steps using the ADAM optimizer with learning rate fixed to $10^{-4}$ on a NVIDIA V100 GPU.
The demo is available 
\href{https://thuhcsi.github.io/interspeech2021-multi-factor-vc/}{https://thuhcsi.github.io/interspeech2021-multi-factor-vc/}.
\subsection{Objective evaluation}
We calculated the Mel-cepstral distortion (MCD) on a sub set of the testing set which consists 543 parallel three-aspects
conversion pairs and the results is shown in Table \ref{tab:MCD}.
The proposed system outperforms the baseline with decreasing the MCD from 3.89 to 3.58.
Here the MCD of the baseline system is calculated based on our own impementation of SpeechSplit \cite{qian2020unsupervised}.

\begin{figure}
\begin{minipage}{0.48\linewidth}
  \centerline{\includegraphics[width=4cm,height=3.6cm]{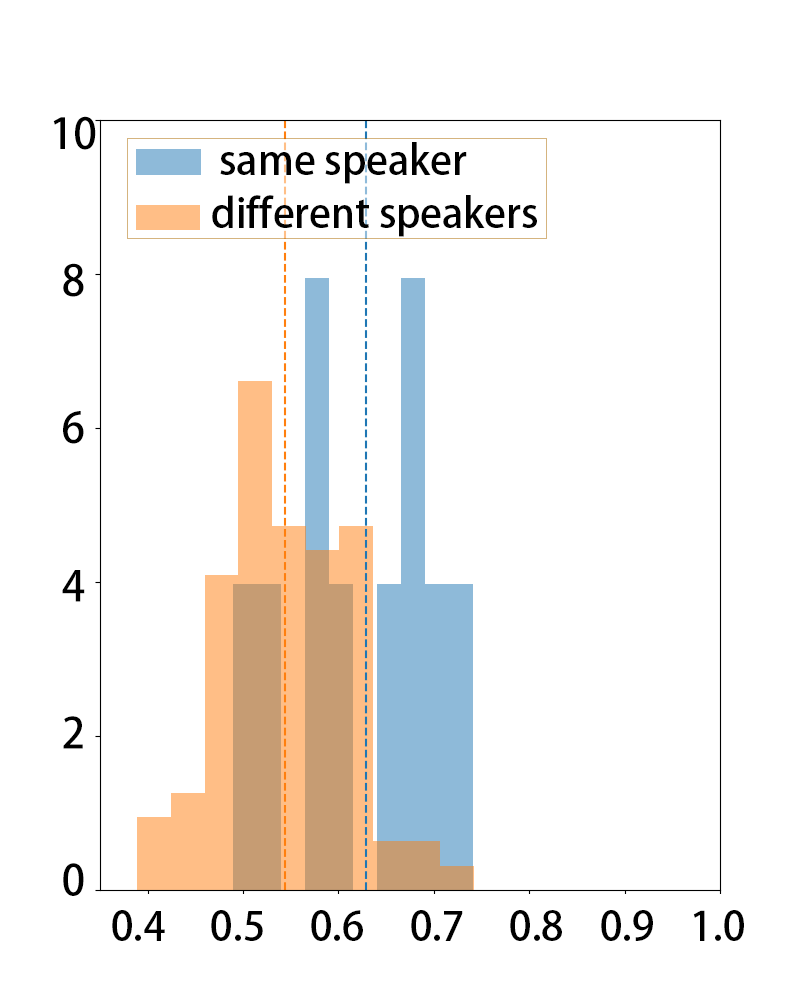}}
  \centerline{(a) \textbf{Baseline} }
\end{minipage}
\begin{minipage}{0.48\linewidth}
  \centerline{\includegraphics[width=4cm,height=3.6cm]{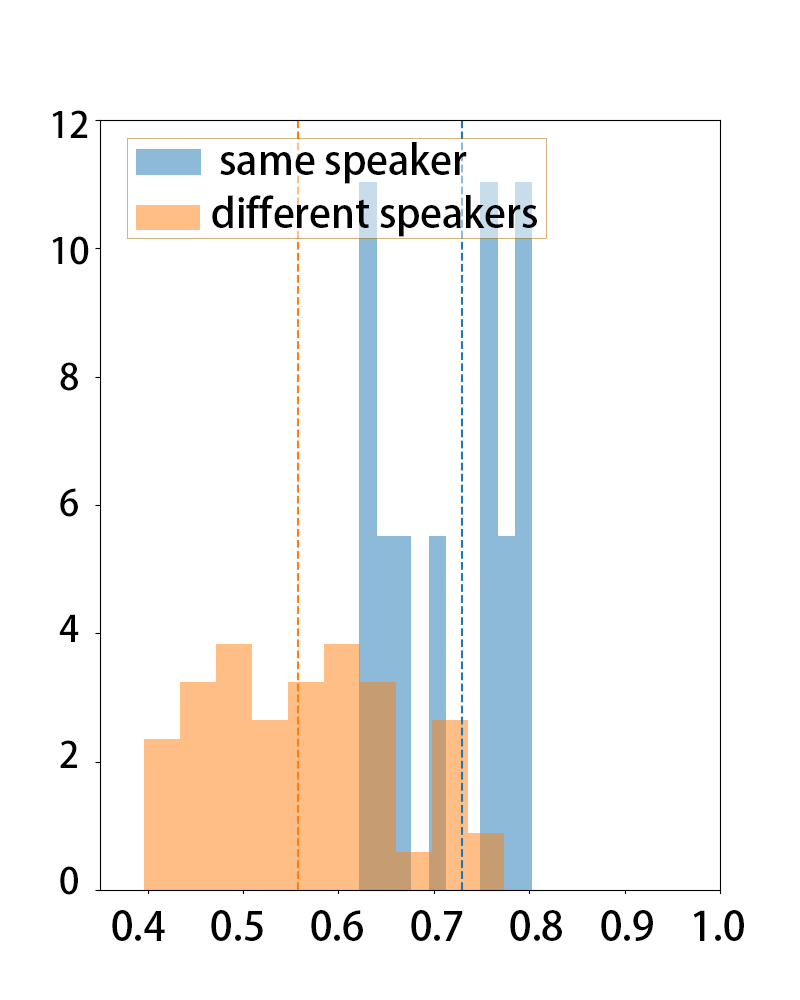}}
  \centerline{(b) \textbf{Proposed} }
\end{minipage}
\caption{\textit{Normalized histogram of similarity values between utterances. The horizontal coordinate denotes the similarity value and the ordinate denotes the number of utterance pairs. The vertical dotted lines denote the median value.
}}
\label{fig:histogram}
\end{figure}

\begin{table}[h]
	\caption{MCD comparison between different systems. 
	}
	\label{tab:MCD}
	\centering
	\begin{tabular}{lcc}
		\toprule
			&\textbf{Baseline} &\textbf{Proposed}\\
		\midrule
		MCD &3.89	&3.58 \\
		\bottomrule
	\end{tabular}
\end{table}
To verify timbre transfer ability of conversion systems, we analysed the speaker confusion results.
There were a number of speakers involved 
and we selected 10 speakers for demonstration.
We randomly selected 10 timbre
conversion involved results of different target speakers.
The reference timbre was extracted from recording audio samples.
We calculated speaker embedding cross-similarity between utterances and the normalized histogram of similarity scores is shown in Figure \ref{fig:histogram}.
Each bar-column represents the number of utterance pairs corresponding to the similarity score. 
For the proposed system,
the maximum similarity score between same speaker is 0.80 and the median score is 0.72.
For
the baseline system,
the maximum similarity score between same speaker is only 0.74.
By comparison, the audios converted by proposed system are more identifiable so as to characterize a specific speaker.
The baseline system has weaker timbre transfer ability as 
the converted audios 
are prone to timbre flipping which is caused by 
speaker information leakage.
However, there exists that similarity scores of different speakers from these two systems exceed 0.6.
We found that when other speech factors 
are converted meanwhile, the performance of speaker classification degrades.

\subsection{Subjective evaluation}
Mean opinion scores (MOS) tests were conducted to evaluate the speech quality, prosody similarity
and timbre similarity.
There were twenty listeners and eight three-aspects conversion pairs for each system involved in the evaluation.
In the prosody similarity MOS, the listeners were asked to rate how similar the converted samples sound to the reference sample in terms of pauses, emphasis and speaking rate.
The evaluation of the three aspects is independent and does not affect each other.
When evaluating the similarity of timbre conversion, for example, we should not pay attention to speech quality or prosody.

\begin{figure}
\begin{minipage}{0.49\linewidth}
  \centerline{\includegraphics[width=4.3cm,height=3.5cm]{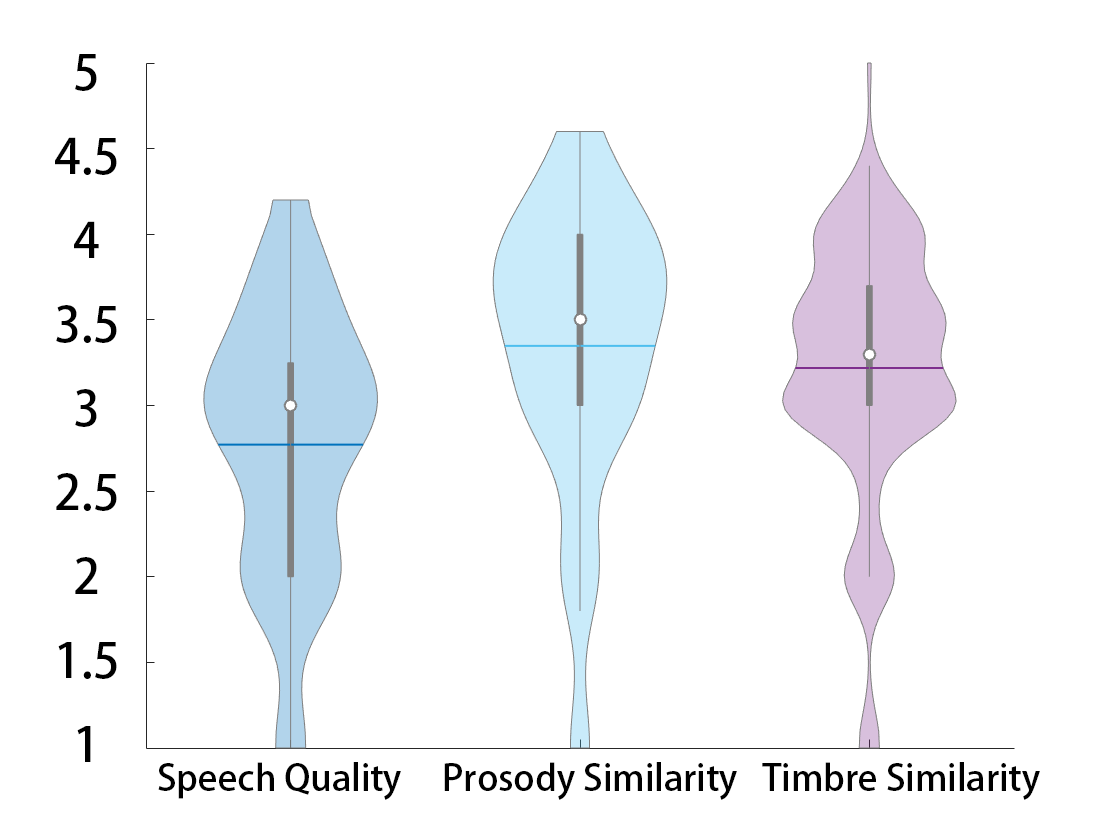}}
  \centerline{(a) \textbf{Baseline} }
\end{minipage}
\begin{minipage}{0.49\linewidth}
  \centerline{\includegraphics[width=4.3cm,height=3.5cm]{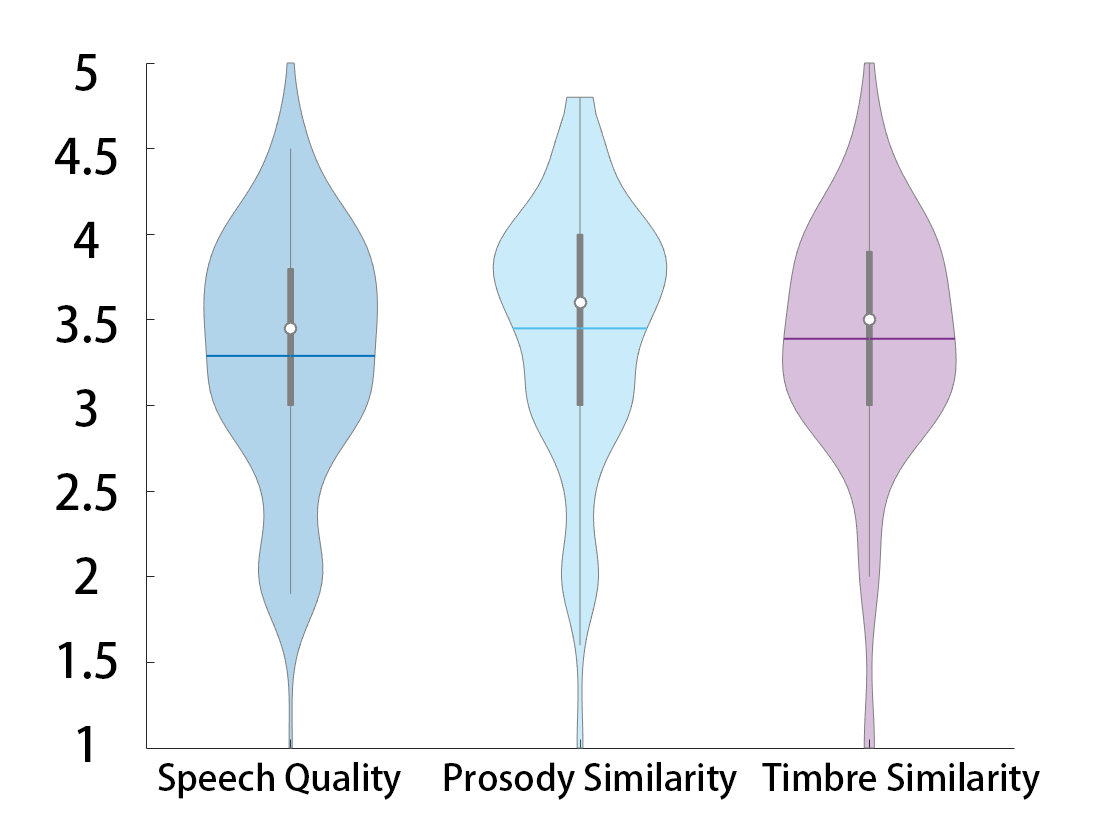}}
  \centerline{(b) \textbf{Proposed} }
\end{minipage}
\caption{\textit{Violin plots of scores obtained in the MOS tests. The horizontal coordinate denotes the three metrics and the ordinate denotes the MOS score.
The horizontal lines 
denote mean value.}}
\label{fig:evaluation}
\end{figure}

The violin plots \cite{hintze1998violin} of results are shown in Figure \ref{fig:evaluation}.
The speech quality of the proposed system is more salient than the baseline.
Most of the quality scores of the proposed system distribute in the range of 3.0$\sim$4.5 with an average of 3.30.
The quality scores of the baseline system mainly distribute in the range of 2.5$\sim$3.5 with an average of 2.79.
The baseline system is prone to generates more low-quality audios as there are quite a little scores are in the range of 1$\sim$1.5.
In terms of prosody transfer, the scores of the two systems have a peak around 3.8 which means the two systems can effectively transfer the prosody.
Nevertheless, the upper limit of the proposed system is higher than the baseline which means that the proposed system can yield higher prosody similarity conversion results.
In terms of timbre transfer,
the baseline system is more likely to produce low-similarity audios by contrast.
Overall, the performance of prosody transfer is better than the timbre transfer.





\subsection{Ablation study}
In the experiment, we observed that while conducting rhythm-only conversion, 
the content from the target-rhythm audio is also encoded into the converted audio.
Target content leaks into rhythm representation
causes the messy 
source and target content mixture.
To evaluate the effect of MBV bottleneck on reducing the amount of content information encoded into the rhythm representation, we calculated the word error rate (WER) on rhythm-only conversion results.
There were 20 non parallel conversion pairs involved and the results are shown in Table \ref{tab:ASR}.
The WER decreases from 43.92\% to 28.89\% after applying the MBV.
The content is wiped out from the rhythm representation more thoroughly and less target content leaks into converted speech.

\begin{table}[]
\caption{\textit{WER on converted speech while performing rhythm-only conversion. 
The `Baseline w D-R' denotes the system that is same as the baseline except MBV is adopted as rhythm code. 
}}
    \centering
\begin{tabular}{lll}
\hline  &\textbf{Baseline} &\textbf{Baseline w D-R}\\
\hline WER(\%) &43.92 &28.89\\
\hline
\end{tabular} 
    \label{tab:ASR}
\end{table}

\begin{figure}[t]
\begin{minipage}{0.48\linewidth}
  \centerline{\includegraphics[width=3cm,height=1.8cm]{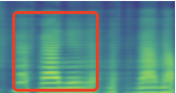}}
  \centerline{(a) \textbf{Baseline} }
\end{minipage}
\begin{minipage}{0.48\linewidth}
  \centerline{\includegraphics[width=3cm,height=1.8cm]{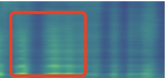}}
  \centerline{(b) \textbf{Proposed} }
\end{minipage}
\caption{\textit{Reconstructed Mel spectrograms when content is removed of the sentence ``I must do something about it."}}
\label{fig:removed}
\end{figure}

To further elucidate the disentanglement performance of our proposed framework, 
we generate mel spectrogram with one component removed by set the corresponding input as zero \cite{qian2020unsupervised}.
As shown in Figure \ref{fig:removed}, after the content information is removed, the spectrogram of the proposed system is composed of 
more uninformative slots and the formant pattern is blur which indicates the missing phone information.
It can be observed that our proposed system removes the content information more thoroughly than the baseline which means that in our system, the amount of content information leaks into other encoder is less.
Given the space limit, Figure \ref{fig:removed} only shows the results of content removed.
The results of other component removed are similar to content which the corresponding right information is missing in the synthesized mel spectrogram.
When the rhythm information is removed, the output spectrogram is blank.
When timbre is removed, the formant position is more random and when pitch is removed, the pitch contour is flatter.

\section{Conclusion}
\label{sec:majhead}

In order to achieve a
highly controllable 
style transfer on multiple factors in VC,
we propose a disentangled speech representation learning framework based on adversarial learning.
We extract four speech representations
and employ MAP network 
to further disentangle speech representations.
Experimental results show that the proposed speech representation learning framework significantly improves the robustness of VC
on multiple factors.
Investigations of the design of masking strategies is left for future work.
\section{Acknowledgement}
This work was conducted when the first author was an intern at Huya Inc., and was supported by National Natural Science Foundation of China (NSFC) (62076144), joint research fund of NSFC-RGC (Research Grant Council of Hong Kong) (61531166002, N\_CUHK404/15), Major Project of National Social Science Foundation of China (NSSF) (13\&ZD189).




\bibliographystyle{IEEEtran}
\bibliography{references}

\end{document}